\documentclass[aps,prb,eqsecnum]{revtex4}
\usepackage{epsfig}
\newcommand{\diag}{{\rm diag\,}}
\newcommand{\tr}{{\rm tr\,}}

\newcommand{\trg}{{\rm trg\,}}
\newcommand{\detg}{{\rm detg\,}}

\begin{document}

\title{Derivation of the Supersymmetric
       Harish--Chandra Integral
       for ${\rm UOSp}(k_1/2k_2)$}

\author{Thomas Guhr$^1$ and Heiner Kohler$^2$}

\affiliation{$^1$ Matematisk Fysik, LTH, Lunds Universitet,
                  Box 118, 22100 Lund, Sweden\\
             $^2$ Departamento de Teor\'{i}a de Materia
                   Condensada, Universidad
                  Aut\'onoma, Madrid, Spain}

\date{\today}

\begin{abstract}
The previous supersymmetric generalization of the unitary
Harish--Chandra integral prompted the conjecture that the
Harish--Chandra formula should extend to all classical supergroups. We
prove this conjecture for the unitary orthosymplectic supergroup ${\rm
UOSp}(k_1/2k_2)$.  To this end, we construct and solve an eigenvalue
equation.
\end{abstract}

\maketitle

\section{Introduction}
\label{sec0}

Harish--Chandra~\cite{HC1} gave a closed formula for a class of group
integrals. Let $\mathcal{G}$ be a compact semi--simple Lie group and
let $a$ and $b$ fixed elements in the Cartan subalgebra
$\mathcal{H}_0$ of $\mathcal{G}$. Harish--Chandra's formula then reads
\begin{equation}
\int\limits_{U\in\mathcal{G}}
        \exp\left(\tr U^{-1}a Ub\right) d\mu(U)
            = \frac{1}{|\mathcal{W}|}\sum_{w\in\mathcal{W}}
              \frac{\exp\left(\tr w(a)b\right)}
                   {\Pi(a)\Pi(w(b))} \ ,
\label{hcf}
\end{equation}
where $d\mu(U)$ stands for the properly normalized invariant measure
and $\Pi(a)$ for the product of all positive roots of
$\mathcal{H}_0$. Moreover, $\mathcal{W}$ is the Weyl reflection group
of $\mathcal{G}$ and $|\mathcal{W}|$ is the number of its elements. We
notice that the integrals~(\ref{hcf}) should not be confused with
Gelfand's spherical functions~\cite{GEL1,HEL,HC,OP}. They are defined
by a group integral which looks at first sight just like the one in
Eq.~(\ref{hcf}), however, for Gelfand's spherical functions, $a$ and
$b$ are not in the Cartan subalgebra. Thus, Gelfand's spherical
functions are very different objects. Only in the case of
$\mathcal{G}={\rm SU}(N)$, the Harish--Chandra integral~(\ref{hcf})
coincides with the unitary spherical function of Gelfand. It is known
as the Itzykson--Zuber integral~\cite{IZ}. A very handy diffusion
equation method was developed in Ref.~\onlinecite{IZ} to derive this
unitary case.

A supersymmetric generalization of the Itzykson--Zuber integral,
i.e.~the extension of the Harish--Chandra integral to the case of the
unitary supergroup ${\rm U}(k_1/k_2)$, was first obtained in
Ref.~\onlinecite{TG} by generalizing the Itzykson--Zuber diffusion
equation method to supersymmetry. In its most general form, this
integral was obtained in Ref.~\onlinecite{GGT} as an application of
Gelfand--Tzetlin coordinates for ${\rm U}(k_1/k_2)$ and also in
Ref.~\onlinecite{AMU} by employing the methods as given in
Ref.~\onlinecite{TG}.

Serganova~\cite{Vera} and Zirnbauer~\cite{MRZ} conjectured that
Harish--Chandra's formula should not only have a supersymmetric
extension in the unitary case, but also generalize to all classical
supergroups $\mathcal{SG}$.  Thus, one would expect a result of the
following form to hold,
\begin{equation}
\int\limits_{u\in\mathcal{SG}} \exp\left(\trg u^{-1}s ur\right) d\mu(u)
            = \frac{1}{|\mathcal{SW}|} \sum_{w\in\mathcal{SW}}
              \frac{\exp\left(\trg w(s)r\right)}
                   {\pi(s)\pi(w(r))} \ ,
\label{shcf}
\end{equation}
where $s$ and $r$ are in the Cartan subalgebra $\mathcal{SH}_0$ of
$\mathcal{SG}$.  The Weyl reflection group $\mathcal{SW}$ and the root
system $\pi(s)$ have to be properly generalized to superspace,
$|\mathcal{SW}|$ is the number of elements in $\mathcal{SW}$. The
extension of the Harish--Chandra integral to the case of the unitary
supergroup ${\rm U}(k_1/k_2)$ is certainly of the
form~(\ref{shcf}). The most interesting remaining case is the unitary
orthosymplectic supergroup ${\rm UOSp}(k_1/2k_2)$. For this case, we
present a proof of the conjecture~(\ref{shcf}) in this note.  In
Sec.~\ref{sec2}, we state and derive the supersymmetric
Harish--Chandra integral for ${\rm UOSp}(k_1/2k_2)$. We summarize and
conclude in Sec.~\ref{sec5}.

\section{The supergroup integral and its derivation}
\label{sec2}

After briefly summarizing properties of the super group ${\rm
UOSp}(k_1/2k_2)$ and introducing our notation in Sec.~\ref{sec21}, we
state the group integral in Sec.~\ref{sec22}. The solution is sketched
in the following two sections. In Sec.~\ref{sec23}, we formulate an
eigenvalue equation for the integral over the supergroup algebra,
which is solved by separation in Sec.~\ref{sec24}.

\subsection{The supergroup ${\rm UOSp}(k_1/2k_2)$}
\label{sec21}

Kac~\cite{KAC1,KAC2} gave a classification of the classical
superalgebras similar to Cartan's classification of the Lie algebras
in ordinary space. In principle, to each classical superalgebra a
supergroup is associated by the exponential mapping.  However, the
classification pattern of the supergroups is usually somewhat
coarser~\cite{RIT}. If one omits the supergroup stemming from the
exceptional superalgebras, one is left with only four different types
of subgroups of the general linear supergroup ${\rm GL}(k_1/k_2)$,
namely the unitary supergroup ${\rm U}(k_1/k_2)$, the orthosymplectic
supergroup ${\rm OSp}(k_1/2k_2)$ whose compact form is denoted by
${\rm UOSp}(k_1/2k_2)$ and the groups associated with the strange
superalgebras ${\rm P}(k)$ and ${\rm Q}(k)$. In most applications, the
${\rm U}(k_1/k_2)$ and the ${\rm UOSp}(k_1/2k_2)$ are used. The ${\rm
OSp}(k_1/2k_2)$ is formed by the elements $u$ of ${\rm GL}(k_1/2k_2)$
which leave invariant the metric
\begin{equation}
L = \diag(1_{k_1},J) \ ,
\label{metL}
\end{equation}
such that $u^T Lu=L$. Here, $J$ is the symplectic metric
\begin{equation}
J = \tau^{(1)}\otimes 1_{k_2}
  = \left[ \begin{array}{cc}
                         0       &  1_{k_2} \\
                         -1_{k_2}       &  0
                         \end{array} \right]
 \ , \qquad {\rm with} \qquad
 \tau^{(1)}=\left[ \begin{array}{cc}
                        0       &  1\\
                        -1      &  0
                        \end{array} \right] \ .
\label{metJ}
\end{equation}
Restricting ${\rm OSp}(k_1/2k_2)$ to its compact part, we arrive at ${\rm
UOSp}(k_1/2k_2)$.  While the fermionic dimension $2k_2$ is always even, the
bosonic dimension $k_1$ can be even or odd, eventually resulting in some slight
differences for the group integral. The superalgebra ${\rm uosp}(k_1/2k_2)$ and
the supergroup ${\rm UOSp}(k_1/2k_2)$ are, as usual, connected via the exponential
mapping. For $\sigma\in {\rm uosp}(k_1/2k_2)$ we have $u=\exp(\sigma) \in {\rm
UOSp}(k_1/2k_2)$. These generators span the algebra. A compact real form of this
algebra can be written as
\begin{equation}
\sigma = \left[ \begin{array}{cc}
                \sigma^{({\rm o})} & \sigma^{({\rm a})} \\
                \sigma^{({\rm a})\dagger} & i\sigma^{({\rm usp})}
                \end{array} \right] \ .
\label{alg}
\end{equation}
One uses the supertrace denoted by $\trg$ as Killing--Cartan form. The $k_1\times
k_1$ matrix $\sigma^{({\rm o})}$ is antisymmetric, it is in the algebra ${\rm
o}(k_1)$ and generates the ordinary group ${\rm O}(k_1)$. The $2k_2\times 2k_2$
matrix $\sigma^{({\rm usp})}$ is in the algebra ${\rm usp}(2k_2)$ and generates
the ordinary group ${\rm USp}(2k_2)$, i.e.~in the basis defined by
Eq.~(\ref{metL}) it is of the form
\begin{equation}
\sigma^{({\rm usp})} = \left[ \begin{array}{cc}
         \sigma^{({\rm usp,1})} & -\sigma^{({\rm usp,2})} \\
         \sigma^{({\rm usp,2})\dagger} & -\sigma^{({\rm usp,1})T}
         \end{array} \right] \ .
\label{algusp}
\end{equation}
Here $\sigma^{({\rm usp,1})}$ is a skew--hermitean matrix and
$\sigma^{({\rm usp,2})}$ is skew--symmetric.  The $k_1\times 2k_2$
matrix $\sigma^{({\rm a})}$ in Eq.~(\ref{alg}) contains the
anticommuting variables, it is in the sector ${\rm
usop}(k_1/2k_2)-{\rm o}(k_1)-{\rm usp}(2k_2)$ and has the symmetry
$\sigma^{({\rm a})*}J=-\sigma^{({\rm a})}$. The asterix denotes the
complex conjugate of the second kind for Grassmann variables.

Particularly important in the present context is the Cartan subalgebra
${\rm uosp}^{(0)}(k_1/2k_2)$ of ${\rm uosp}(k_1/2k_2)$. As in the
theory of Lie--algebrae in ordinary space there is a difference for
the orthogonal group in even or odd dimension. We introduce the
notation $[k_1/2]$ for the integer part of $k_1/2$. Then, for even
bosonic dimension $[k_1/2]=k_1/2$ , the elements of ${\rm
uosp}^{(0)}(k_1/2k_2)$ are the matrices
\begin{equation}
s = \diag(is_{11}\tau^{(1)},\ldots,is_{[k_1/2]1}\tau^{(1)},
          is_{12},\ldots,is_{k_21},-is_{11},\ldots,-is_{k_21}) \ ,
\label{csue}
\end{equation}
while for odd bosonic dimension $[k_1/2]= (k_1-1)/2$, the Cartan subalgebra ${\rm
uosp}^{(0)}(k_1/2k_2)$ consists of the matrices
\begin{equation}
s = \diag(is_{11}\tau^{(1)},\ldots,is_{[k_1/2]1}\tau^{(1)},0,
          is_{12},\ldots,is_{k_21},-is_{11},\ldots,-is_{k_21}) \ .
\label{csuo}
\end{equation}
Thus, ${\rm uosp}^{(0)}(k_1/2k_2)$ is the direct sum of the Cartan
subalgebras of ${\rm o}(k_1)$ and ${\rm usp}(2k_2)$.

\subsection{Statement of the supergroup integral}
\label{sec22}

Using the definitions of the previous section, we can write
formula~(\ref{shcf}) for the case of the supergroup ${\rm
UOSp}(k_1/2k_2)$ more explicitly. For two fixed elements $s$ and $r$
of the Cartan subalgebra ${\rm uosp}^{(0)}(k_1/2k_2)$, we have
\begin{eqnarray}
&&\int\limits_{u\in {\rm UOSp}(k_1/2k_2)}
   \exp\left(i\trg u^{-1}s ur\right) d\mu(u)
\nonumber\\
&& \qquad\qquad\qquad\qquad
     = \frac{1}{[k_1/2]!k_2!}
    \frac{\bigl(\det[\cos(2s_{p1}r_{q1})]+
                \det[i\sin(2s_{p1}r_{q1})]\bigr)
                      \det[-2i\sin(2s_{p2}r_{q2})]}
         {B_{k_12k_2}(s)B_{k_12k_2}(r)}
\label{shcu1}
\end{eqnarray}
for $k_1$ even and
\begin{equation}
\int\limits_{u\in {\rm UOSp}(k_1/2k_2)}
  \exp\left(i\trg u^{-1}s ur\right) d\mu(u)  =\frac{1}{[k_1/2]!k_2!}
    \frac{\det[i\sin(2s_{p1}r_{q1})]\det[2\cos(2s_{p2}r_{q2})]}
         {B_{k_12k_2}(s)B_{k_12k_2}(r)}
\label{shcu2}
\end{equation}
for $k_1$ odd. We introduced the function $B_{k_12k_2}(s)$ which is
given by
\begin{equation}
B_{k_12k_2}(s) = 2^{k_2}\frac{\prod_{p<q}(s_{p1}^2-s_{q1}^2)
                      \prod_{p<q}(s_{p2}^2-s_{q2}^2)
                      \prod_{p=1}^{k_2}s_{p2}}
                     {\prod_{p,q}(s_{p1}^2+s_{q2}^2)}
\label{bke}
\end{equation}
for even bosonic dimension $k_1$ and by
\begin{equation}
B_{k_12k_2}(s) = 2^{k_2}\frac{\prod_{p<q}(s_{p1}^2-s_{q1}^2)
                      \prod_{p<q}(s_{p2}^2-s_{q2}^2)
                      \prod_{p=1}^{[k_1/2]}s_{p1}}
                     {\prod_{p,q}(s_{p1}^2+s_{q2}^2)}
\label{bko}
\end{equation}
for odd bosonic dimension $k_1$. These two formulae differ only in the
last terms of the numerators.

Formulae~(\ref{shcu1}) and (\ref{shcu2}) contain, as special cases,
the ordinary orthogonal and unitary symplectic Harish--Chandra
integrals for $\mathcal{G}={\rm SO}(k_1)$ and for $\mathcal{G}={\rm
USp}(2k_2)$, if we set $k_2=0$ or $k_1=0$, respectively. Thus, the
derivation of the supersymmetric integral to follow also includes a
rederivation of those ordinary integrals. For equal bosonic and
fermionic dimension, formula~(\ref{shcu1}) was conjectured in
Ref.~\onlinecite{MRZ} and used to calculate the correlation functions
in a certain circular random matrix ensemble.

We mention in passing that the invariant measure $d\mu(u)$ and its
normalization relate to non--trivial questions of certain boundary
contributions in superanalysis~\cite{ROT} which are highly important
in applications. In the present context, however, we do not need to go
into this.

\subsection{Eigenvalue equation}
\label{sec23}

The main idea for the derivation of formulae~(\ref{shcu1}) and
(\ref{shcu2}) is to properly modify the supersymmetric
extension~\cite{TG} of the Itzykson--Zuber diffusion equation
method~\cite{IZ} to the present case.  It turns out that it is
somewhat more convenient to construct the eigenvalue equation
associated with the diffusion equation. The two equations are related
by Fourier expansion. Such an eigenvalue equation for the ordinary
case of ${\rm SU}(N)$ as originally discussed by Itzykson and
Zuber~\cite{IZ} was constructed by Br\'ezin~\cite{BREZ}. Berezin and
Karpelevich~\cite{BK} had studied such an eigenvalue equation to
calculate the twofold group integral named after them, see also
Ref.~\onlinecite{HOG}. To construct the eigenvalue equation needed to
derive formulae~(\ref{shcu1}) and (\ref{shcu2}), we adjust the steps
made in Refs.~\onlinecite{GUWE,JSV}, where a supersymmetric eigenvalue
equation was employed for the extension of the Berezin--Karpelevich
integral.

We introduce the Laplace operator over the superalgebra ${\rm
uosp}(k_1/2k_2)$
\begin{equation}
\Delta =\frac{1}{2}\sum_{p<q}^{k_1}
        \frac{\partial^2}
        {\partial\sigma^{({\rm o})2}_{pq}} +\sum_{p,q}^{2k_2}
        \frac{1+\delta_{pq}}{4}
\frac{\partial^2}{\partial\sigma^{({\rm usp})}_{pq}
\partial\sigma^{({\rm usp})*}_{pq}}+
\frac{1}{2}\sum_{p=1}^{k_1}\sum_{q=1}^{k_2}
\frac{\partial^2}{\partial\sigma^{({\rm a})}_{pq}
\partial\sigma^{({\rm a})*}_{pq}} \ .
\label{lapc}
\end{equation}
Its eigenfunctions are the plane waves $\exp(i\trg\sigma\rho)$ with both matrices
$\sigma,\rho\in {\rm uosp}(k_1/2k_2)$. Thus, we have
\begin{equation}
\Delta \exp(i\trg\sigma\rho) \ = \ -\trg\rho^2
\exp(i\trg\sigma\rho) \ .
\label{ev}
\end{equation}
We now diagonalize both matrices according to $\sigma=u^{-1}su$ and
$\rho=v^{-1}rv$ where $u$ and $v$ are in the supergroup
${\rm UOSp}(k_1/2k_2)$ and $s$ and $r$ are in the Cartan subalgebra
${\rm uosp}^{(0)}(k_1/2k_2)$, i.e.~given by Eq.~(\ref{csue}) or
Eq.~(\ref{csuo}), respectively. Integrating both sides over $v$
and using the invariance of the measure $d\mu(v)$, we arrive at
the radial eigenvalue equation
\begin{equation}
\Delta_s \chi_{k_12k_2}(s,r) =
                  -\trg r^2\chi_{k_12k_2}(s,r) \ ,
\label{evr}
\end{equation}
where, now using $u$ instead of $v$ again, the function
\begin{equation}
\chi_{k_12k_2}(s,r) = \int\limits_{u\in {\rm UOSp}(k_1/2k_2)}
                       \exp\left(i\trg u^{-1}s ur\right) d\mu(u)
\label{chi}
\end{equation}
is the integral we want to calculate. The operator $\Delta_s$ in
Eq.~(\ref{evr}) is the radial part of $\Delta$. The term radial refers
to the Cartan subalgebra. This usage which is common in mathematics
should not lead to confusions with the radial operators used, for
example, in Refs.~\onlinecite{GUKOP1,GUKO1,GUKOP2,GUKO2} where quite
different spaces were studied.  To obtain the radial operator
$\Delta_s$, we need the Jacobian, or Berezinian, of the variable
transformation $\sigma=u^{-1}su$.  This Berezinian is given by the the
functions $B_{k_12k_2}^2(s)$ of Eqs.~(\ref{bke}) and~(\ref{bko}). It
was not possible for us to find out where this Berezinian was first
obtained, and we do not claim originality for its calculation.  In any
case, to make the paper self--contained, we sketch the calculation in
Appendix~\ref{appA}. Hence, the radial part of the Laplacean over
${\rm uosp}(k_1/2k_2)$ reads
\begin{equation}
\Delta_s = \frac{1}{2}\sum_{p=1}^{[k_1/2]} \frac{1}{B_{k_12k_2}^2(s)}
           \frac{\partial}{\partial s_{p1}}B_{k_12k_2}^2(s)
           \frac{\partial}{\partial s_{p1}} +
           \frac{1}{2}\sum_{p=1}^{k_2} \frac{1}{B_{k_12k_2}^2(s)}
           \frac{\partial}{\partial s_{p2}}B_{k_12k_2}^2(s)
           \frac{\partial}{\partial s_{p2}} \ .
\label{lapr}
\end{equation}
The number of bosonic eigenvalues is $[k_1/2]$, i.e.~identical for the
pairs ${\rm uosp}(k_1/2k_2)$ and ${\rm uosp}(k_1+1/2k_2)$ with $k_1$
even. However, the operator $\Delta_s$ is not the same in these two
cases, because the functions~(\ref{bke}) and~(\ref{bko}) differ.

\subsection{Solution by separation}
\label{sec24}

The Laplacean~(\ref{lapr}) is separable. We make an ansatz for the
group integral which separates off the square roots of the Berezinians,
\begin{equation}
\chi_{k_12k_2}(s,r) = \frac{\omega_{k_12k_2}(s,r)}
                           {B_{k_12k_2}(s)B_{k_12k_2}(r)} \ .
\label{ans}
\end{equation}
A tedious but straightforward calculation then yields the trivial
eigenvalue equation
\begin{equation}
\frac{\partial^2}{\partial\vec{s}^{\, 2}}
\omega_{k_12k_2}(s,r) = -\trg r^2\omega_{k_12k_2}(s,r) \ .
\label{evo}
\end{equation}
for the function $\omega_{k_12k_2}^{(\beta)}(s,r)$. Here we introduced
the gradient
\begin{equation}
\frac{\partial}{\partial\vec{s}} =
\left(\frac{\partial}{\partial s_{11}},\ldots,
      \frac{\partial}{\partial s_{[k_1/2]1}},
      \frac{\partial}{\partial s_{12}},\ldots,
      \frac{\partial}{\partial s_{k_22}}\right) \ ,
\label{grad}
\end{equation}
which also defines the flat Laplacean $\partial^2/\partial\vec{s}^{\,
2}$ appearing in the eigenvalue equation~(\ref{evo}). A crucial
feature of the square root of the Berezinian enters the derivation of
Eq.~(\ref{evo}). It satisfies the harmonic equation
\begin{equation}
\frac{\partial^2}{\partial\vec{s}^{\, 2}} B_{k_12k_2}(s) = 0 \ ,
\label{harm}
\end{equation}
which we prove in Appendix~\ref{appB}. Any linear combination of
products of exponentials solves the eigenvalue
equation~(\ref{evo}). However, as the group integral and the
eigenvalue equations are obviously invariant under permutations of the
variables in the ${\rm o}(k_1)$ sector $s_{p1}, \ p=1,\ldots,[k_1/2]$
or, equivalently, $r_{p1}, \ p=1,\ldots,[k_1/2]$ and under
permutations of the variables in the ${\rm uosp}(2k_2)$ sector
$is_{p2}, \ p=1,\ldots,k_2$ or, equivalently, $ir_{p2}, \
p=1,\ldots,k_2$, the desired solution must have the same
property. Moreover, there is a symmetry under a parity transformation
for the variables $is_{p2}, \ p=1,\ldots,k_2$ and $ir_{p2}, \
p=1,\ldots,k_2$. That is, the solution must be invariant under the
substitution $s_{p2}\rightarrow -s_{p2}$ and $r_{p2}\rightarrow
-r_{p2}$. For $k_1$ odd, the same symmetry must hold also for $s_{p1},
\ p=1,\ldots,[k_1/2]$ and $r_{p1}, \ p=1,\ldots,[k_1/2]$
respectively. By these symmetries the solution of Eq.~(\ref{evo}) for
$k_1$ odd is up to normalization uniquely determined,
\begin{equation}
\omega_{k_12k_2}(s,r)=\frac{1}{[k_1/2]!k_2!}
                      \det[i\sin(2s_{p1}r_{q1})]
                      \det[2\cos(2s_{p2}r_{q2})] \ .
\label{omsu}
\end{equation}
For $k_1$ even, the part antisymmetric under the parity
transformation has to be kept and we find
\begin{equation}
\omega_{k_12k_2}(s,r)=\frac{1}{[k_1/2]!k_2!}
                      \bigl(\det[\cos(2s_{p1}r_{q1})]+
                      \det[i\sin(2s_{p1}r_{q1})]\bigr)
                      \det[-2i\sin(2s_{p2}r_{q2})] \ .
\label{omsg}
\end{equation}
These results give, together with the ansatz~(\ref{ans}), the desired
group integrals~(\ref{shcu1}) and~(\ref{shcu2}).  As already
mentioned, our derivation of the supersymmetric group integral
contains as special cases a rederivation of the ordinary orthogonal
and unitary symplectic Harish--Chandra integrals for $k_2=0$ or
$k_1=0$, respectively.

\section{Summary and conclusions}
\label{sec5}

We calculated the supersymmetric Harish--Chandra integral for the
unitary orthosymplectic supergroup, thereby proving a
conjecture~\cite{Vera,MRZ}. Our derivation uses a diffusion equation
or, equivalentely, eigenvalue equation method. It is based on the
separability of the Laplacean.  Our present contribution is a further
extension of this technique, which, to the best of our knowledge, has
previously only been used for group integrals over the unitary group:
orginally, it was introduced for the Harish--Chandra integral over the
ordinary unitary group~\cite{IZ,BREZ}, then extended for the
supersymmetric Harish--Chandra integral over the unitary
supergroup~\cite{TG}.  Already in 1958, Berezin and
Karpelevich~\cite{BK} had developed such a technique for an integral
over two unitary groups, see also Ref.~\onlinecite{HOG}. This was also
extended to the supersymmetric case~\cite{GUWE,JSV}. Here, we
considered the unitary orthosymplectic supergroup and adjusted the
eigenvalue equation method to this case. As the unitary
orthosymplectic supergroup contains the ordinary orthogonal and
unitary symplectic groups as subgroups and special cases, we
automatically also extended the eigenvalue equation method to these
two ordinary groups.

We are aware of only two methods which could be an alternative:
character expansions and Gelfand--Tzetlin coordinates. Balantekin
developed the character expansion method for the unitary ordinary and
supergroup~\cite{BAL1,BAL2} and obtained various group integrals.
Recently, this method was further extended and employed in
Ref.~\onlinecite{SW}. Similar considerations are also of interest if
one studies the Itzykson--Zuber integral for matrices of large
dimension~\cite{ZJZ}.  Moreover, character expansions could also be
developed for the calculation of certain integrals over the ordinary
orthogonal and unitary symplectic group~\cite{BAL3}, but
Harish--Chandra integrals have so far not been tackled with this
approach. Gelfand--Tzetlin coordinates~\cite{GT,BR} allow one to
compute the ordinary~\cite{SLS} and supersymmetric~\cite{GGT}
Itzykson--Zuber integral directly, i.e.~without using a diffusion or
eigenvalue equation.  This method has not been applied yet to work
out Harish--Chandra integrals for the ordinary orthogonal or unitary
symplectic or the supersymmetric unitary orthosymplectic
supergroup. However, it has been employed for Gelfand's spherical
functions~\cite{GUKOP1,GUKO1,GUKOP2,GUKO2}.

Considering all the cases, in which non--trivial group integrals could
be obtained for the first time or in which known results could be
rederived faster, the diffusion or eigenvalue equation method used and
further extended here shows a remarkably wide range of applicability.

\begin{acknowledgments}
TG and HK acknowledge financial support from the Swedish Research
Council and from the RNT Network of the European Union with Grant
No.~HPRN--CT--2000-00144, respectively. HK also thanks the division of
Mathematical Physics, LTH, for its hospitality during his visits to
Lund.
\end{acknowledgments}

\appendix

\section{Calculation of the Berezinian}
\label{appA}

We use the standard procedure of obtaining the metric tensor $g$ whose
superdeterminant is the square of the Berezinian. The variation of the
element $\sigma=u^{-1}su \in {\rm uosp}(k_1/2k_2)$ reads
\begin{equation}
d\sigma=u^{-1}(ds+[s,\delta\widetilde{u}])u \ ,
\quad {\rm where} \qquad \delta\widetilde{u}=duu^{-1}
\label{vari}
\end{equation}
is also in the algebra ${\rm uosp}(k_1/2k_2)$. Thus, the invariant
length element is given by
\begin{eqnarray}
\trg d\sigma^2 &=& \trg(ds+[s,\delta\widetilde{u}])^2
                =  \trg ds^2 + \trg[s,\delta\widetilde{u}]^2
                                           \nonumber\\
               &=& \sum_{p=1}^{[k_1/2]} ds_{p1}^2 +
                   \sum_{p=1}^{k_2} ds_{p2}^2 +
                   \sum_n (\alpha_n^{({\rm o})})^2
                          (\delta\widetilde{u}_n^{({\rm o})})^2 +
                   \sum_n (\alpha_n^{({\rm usp})})^2
                          (\delta\widetilde{u}_n^{({\rm usp})})^2 +
                   \sum_n (\alpha_n^{({\rm a})})^2
                          (\delta\widetilde{u}_n^{({\rm a})})^2 \ .
\label{invl}
\end{eqnarray}
In the last step, we expanded the traces, the metric $g$ can then be
read of from the coefficients in front of the squared variation
differentials. We split the contribution from the commutator in three
terms and introduced a new index $n$, labelling the roots and the
variation differentials stemming from $\delta\widetilde{u}$. There are
three types of roots, corresponding to the ${\rm o}(k_1)$ and the
${\rm usp}(2k_2)$ subalgebras and the remaining sector ${\rm
uosp}(k_1/2k_2)-{\rm o}(k_1)-{\rm usp}(2k_2)$ containing the
anticommuting degrees of freedom. For even bosonic dimension $2k_1$,
there are $2k_1(k_1-1)$ roots $\alpha_n^{({\rm o})}$ of ${\rm
o}(2k_1)$, given by $\pm s_{p1} \pm s_{q1}$ with independent signs and
$p<q$. For odd bosonic dimension $2k_1+1$, there are $2k_1$ additional
roots $\pm s_{p1}$ which are needed for the complete root system of
${\rm o}(2k_1+1)$.  The roots $\alpha_n^{({\rm usp})}$ of ${\rm
usp}(2k_2)$ are $\pm is_{p2} \pm is_{q2}$ with independent signs and
$p<q$, and furthermore $\pm 2is_{p2}$, all together $2k_2^2$
roots. Finally, we have the $2k_1k_2$ roots $\alpha_n^{({\rm a})}$
from ${\rm uosp}(k_1/2k_2)-{\rm o}(k_1)-{\rm usp}(2k_2)$, which read
$\pm s_{p1} \pm is_{q2}$ with independent signs and indices $p,q$. For
odd bosonic dimension we have $2k_1$ additional roots $\alpha_n^{({\rm
a})}=\pm is_{p2}$. Collecting everything, the superdeterminant $\detg
g$ of the metric $g$ is the product of all roots from ${\rm o}(k_1)$
and ${\rm usp}(2k_2)$, divided by the product of all roots from ${\rm
uosp}(k_1/2k_2)-{\rm o}(k_1)-{\rm usp}(2k_2)$.  The square root of
$\detg g$ then gives the Berezinians~(\ref{bke}) and~(\ref{bko}).

\section{Square root of the Berezinian as a harmonic function}
\label{appB}

The result~(\ref{harm}) is crucial for the separation ansatz and for
the derivation of the ensuing eigenvalue equations. It is tedious, but
elementary to prove it by explicit calculation. We consider even
$k_1$, the case of odd $k_1$ is treated in the same way. Using
relations such as
\begin{equation}
\sum_{p\neq q} \frac{1}{s_{p1}^2-s_{q1}^2} = 0
\qquad {\rm and} \qquad
\sum_{p\neq q \neq t} \frac{s_{p1}^2}
        {(s_{p1}^2-s_{q1}^2)(s_{p1}^2-s_{t1}^2)} = 0
\label{derbrel}
\end{equation}
we find
\begin{equation}
\frac{1}{B_{k_12k_2}(s)}
\sum_{p=1}^{[k_1/2]}
    \frac{\partial^2}{\partial s_{p1}^2}B_{k_12k_2}(s) =
\sum_{p,q,t} \frac{4s_{p1}^2}
        {(s_{p1}^2+s_{q2}^2)(s_{p1}^2+s_{t2}^2)}
-\sum_{p\neq q, t} \frac{8s_{p1}^2}
        {(s_{p1}^2-s_{q1}^2)(s_{p1}^2+s_{t2}^2)}
-\sum_{p,q} \frac{2}
        {(s_{p1}^2+s_{q2}^2)} \ .
\label{derb}
\end{equation}
Similarly, we obtain
\begin{equation}
\frac{1}{B_{k_12k_2}(s)}
\sum_{p=1}^{k_2}
    \frac{\partial^2}{\partial s_{p2}^2}B_{k_12k_2}(s) =
\sum_{p,q,t} \frac{4s_{p2}^2}
        {(s_{p1}^2+s_{q2}^2)(s_{p1}^2+s_{t2}^2)}
-\sum_{p\neq q, t} \frac{8s_{p2}^2}
        {(s_{p2}^2-s_{q2}^2)(s_{p1}^2+s_{t2}^2)}
-\sum_{p,q} \frac{6}{s_{p1}^2+s_{q2}^2} \ .
\label{derf}
\end{equation}
Combining these two intermediate results, we arrive at
\begin{eqnarray}
\frac{1}{B_{k_12k_2}(s)}
\frac{\partial^2}{\partial\vec{s}^{\, 2}}B_{k_12k_2}(s) &=&
\sum_{p\neq q, t} \left(\frac{4s_{t2}^2}
        {(s_{p1}^2+s_{t2}^2)(s_{q1}^2+s_{t2}^2)}
                       - \frac{8s_{p1}^2}
        {(s_{p1}^2-s_{q1}^2)(s_{p1}^2+s_{t2}^2)}\right) +
                        \nonumber\\
& & \sum_{p\neq q, t} \left(\frac{4s_{t1}^2}
        {(s_{t1}^2+s_{p2}^2)(s_{t1}^2+s_{q2}^2)}
                       - \frac{8s_{p2}^2}
        {(s_{p2}^2-s_{q2}^2)(s_{p1}^2+s_{t2}^2)}\right) +
                        \nonumber\\
& & \sum_{p,q} \left(\frac{8s_{p1}^2}{(s_{p1}^2+s_{q2}^2)^2}
             +\frac{8s_{q1}^2}{(s_{q1}^2+s_{p2}^2)^2}\right) -
\sum_{p,q}\frac{8}{s_{p1}^2+s_{q2}^2}
                         \nonumber\\
&=& \sum_{p\neq q, t} \left(\frac{4s_{t2}^2}
        {(s_{p1}^2+s_{t2}^2)(s_{q1}^2+s_{t2}^2)} -
                            \frac{4s_{p1}^2}
        {(s_{p1}^2-s_{q1}^2)(s_{p1}^2+s_{t2}^2)} -
                            \frac{4s_{q1}^2}
        {(s_{q1}^2-s_{p1}^2)(s_{q1}^2+s_{t2}^2)}\right) +
                        \nonumber\\
& & \sum_{p\neq q, t} \left(\frac{4s_{t1}^2}
        {(s_{p1}^2+s_{t2}^2)(s_{q1}^2+s_{t2}^2)} -
                            \frac{4s_{p2}^2}
        {(s_{p2}^2-s_{q2}^2)(s_{p2}^2+s_{t1}^2)} -
                            \frac{4s_{q2}^2}
        {(s_{q2}^2-s_{p2}^2)(s_{q2}^2+s_{t1}^2)}\right)
                         \nonumber\\
&=& 0 \ ,
\label{derc}
\end{eqnarray}
which is Eq.~(\ref{harm}).


\begin{thebibliography}{30}

\bibitem{HC1}         Harish-Chandra,
                      Am. J. Math. {\bf 79}, 87 (1957)
\bibitem{GEL1}        I.M. Gelfand,
                      Dokl. Akad. Nauk. SSSR {\bf 70} 5 (1950)
\bibitem{HEL}         S. Helgason,
                      Groups and Geometric Analysis,
                      San Diego: Academic Press, 1984
\bibitem{HC}          Harish-Chandra,
                      Am. J. Math. {\bf 80}, 241 (1958)
\bibitem{OP}          M.A. Olshanetsky and A.M. Perelomov,
                      Phys. Rep. {\bf 94}, 313 (1983)
\bibitem{IZ}          C. Itzykson and J.B. Zuber,
                      J. Math. Phys. {\bf 21}, 411 (1980)
                      Phys. Rep. {\bf 129}, 367 (1985)
\bibitem{TG}          T. Guhr,
                      J. Math. Phys. {\bf 32}, 336 (1991)
\bibitem{GGT}         T. Guhr,
                      Commun. Math. Phys. {\bf 176}, 555 (1996)
\bibitem{AMU}         J. Alfaro, R. Medina and L. Urrutia,
                      J. Math. Phys. {\bf 36}, 3085 (1995)
\bibitem{Vera}        V. Serganova,
                      private communication,
                      Berkeley (1992)
\bibitem{MRZ}         M.R. Zirnbauer,
                      J. Phys. {\bf A29}, 7113 (1996)
\bibitem{KAC1}        V.C. Kac,
                      Comm. Math. Phys. {\bf 53}, 31 (1977)
\bibitem{KAC2}        V.C. Kac,
                      Advances in Math. {\bf 26}, 8 (1977)
\bibitem{RIT}         V. Rittenberg,
                      A Guide to Lie Superalgebras,
                      Lecture Notes in Physics {\bf 79},
                      Berlin: Springer--Verlag, 1977
\bibitem{ROT}         M.J. Rothstein,
                      Trans. Am. Math. Soc. {\bf 299}, 387 (1987)
\bibitem{BREZ}        E. Br\'ezin,
                      in: Two dimensional quantum gravity and
                      random surfaces, p. 37,
                      D.J. Gross, T. Piran and S. Weinberg (eds.),
                      Singapore: World Scientific, 1992
\bibitem{BK}          F.A. Berezin and F.I. Karpelevich,
                      Dokl. Akad. NAUK (SSSR) {\bf 118}, 9 (1958)
\bibitem{HOG}         B. Hoogenboom,
                      Ark. Mat. {\bf 20}, 69 (1982)
\bibitem{GUWE}        T. Guhr and T. Wettig,
                      J. Math. Phys. {\bf 37}, 6395 (1996)
\bibitem{JSV}         A.D. Jackson, M.K. Sener and J.J.M. Verbaarschot,
                      Nucl. Phys. {\bf B506}, 612 (1997)
\bibitem{GUKOP1}      T. Guhr and H. Kohler,
                      {\tt math-ph/0011007}
\bibitem{GUKO1}       T. Guhr and H. Kohler,
                      J. Math. Phys. {\bf 43}, 2707 (2002)
\bibitem{GUKOP2}      T. Guhr and H. Kohler,
                      {\tt math-ph/0012047}
\bibitem{GUKO2}       T. Guhr and H. Kohler,
                      J. Math. Phys. {\bf 43}, 2741 (2002)
\bibitem{BAL1}        A.B. Balantekin,
                      Phys. Rev. {\bf D62}, 085017 (2000);
                      {\tt hep-th/0007161}
\bibitem{BAL2}        A.B. Balantekin,
                      Phys. Rev. {\bf E64}, 066105 (2001);
                      {\tt cond-mat/0109112}
\bibitem{SW}          B. Schlittgen and T. Wettig,
                      {\tt math-ph/0209030}
\bibitem{ZJZ}         P. Zinn--Justin and J.B. Zuber,
                      {\tt math-ph/0209019}
\bibitem{BAL3}        A.B. Balantekin and P. Cassak
                      J. Math. Phys. {\bf 43}, 604 (2002)
\bibitem{GT}          I.M. Gelfand and M.L. Tzetlin,
                      Dokl. Akad. Nauk. {\bf 71}, 825 (1950)
\bibitem{BR}          A.O. Barut and R. Raczka,
                      Theory of Group Representations and Applications,
                      Warszawa: Polish Scientific Publishers, 1980
\bibitem{SLS}         S.L. Shatashvili,
                      Commun. Math. Phys. {\bf 154}, 421 (1993)
\end{thebibliography}
\end{document}